\input harvmac
\input epsf
\input amssym
%
%
\noblackbox
\newcount\figno
\figno=0
\def\fig#1#2#3{
\par\begingroup\parindent=0pt\leftskip=1cm\rightskip=1cm\parindent=0pt
\baselineskip=11pt
\global\advance\figno by 1
\midinsert
\epsfxsize=#3
\centerline{\epsfbox{#2}}
\vskip -21pt
{\bf Fig.\ \the\figno: } #1\par
\endinsert\endgroup\par
}
\def\figlabel#1{\xdef#1{\the\figno}}
\def\encadremath#1{\vbox{\hrule\hbox{\vrule\kern8pt\vbox{\kern8pt
\hbox{$\displaystyle #1$}\kern8pt}
\kern8pt\vrule}\hrule}}

\def\frac#1#2{{#1 \over #2}}

\def\p{\partial}
\def\semi{\subset\kern-1em\times\;}
\def\bar#1{\overline{#1}}
\def\sqr#1#2{{\vcenter{\vbox{\hrule height.#2pt
\hbox{\vrule width.#2pt height#1pt \kern#1pt \vrule width.#2pt}
\hrule height.#2pt}}}}

\def\p{\partial}

\def\ad{\bar a}

\def\p{\partial}

%

%
\def\ZZ{\Bbb{Z}}



\Title{\vbox{\baselineskip12pt
\hbox{HIP-2008-16/TH}
}} {\vbox{\centerline {Quantum Hall Effect in AdS/CFT}}}
\centerline{Esko Keski-Vakkuri\foot{esko.keski-vakkuri@helsinki.fi} and Per Kraus\foot{pkraus@ucla.edu}}

\bigskip
 \centerline{${}^1$\it{Helsinki Institute of Physics and Department of Physics}}\centerline{\it{P.O.Box 64, FIN-00014 University of Helsinki, Finland}} \vskip.2cm\centerline{${}^2$\it{Department of Physics and Astronomy,
UCLA,}}\centerline{\it{ Los Angeles, CA 90095-1547,
USA.}}

\baselineskip15pt

\vskip .3in

\centerline{\bf Abstract}

Drawing on the connection with superconductivity, we give a simple AdS
realization of the quantum Hall effect.   The theory includes a
statistical gauge field with a Chern-Simons term, in analogy with
effective field theory models of the QHE.

\Date{May, 2008}


\lref\HartnollVX{
  S.~A.~Hartnoll, C.~P.~Herzog and G.~T.~Horowitz,
  ``Building an AdS/CFT superconductor,''
  arXiv:0803.3295 [hep-th].
}

\lref\DineUI{
  M.~Dine,
  ``Supersymmetry phenomenology (with a broad brush),''
  arXiv:hep-ph/9612389.
}

\lref\WeinbergUV{
  S.~Weinberg,
  ``Non-renormalization theorems in non-renormalizable theories,''
  Phys.\ Rev.\ Lett.\  {\bf 80}, 3702 (1998)
  [arXiv:hep-th/9803099].
}

\lref\SakamotoCX{
  M.~Sakamoto and H.~Yamashita,
  ``A simple proof of the non-renormalization of the Chern-Simons coupling,''
  Phys.\ Lett.\  B {\bf 476}, 427 (2000)
  [arXiv:hep-th/9910200].
}

\lref\ZhangWY{
  S.~C.~Zhang, T.~H.~Hansson and S.~Kivelson,
  ``An effective field theory model for the fractional quantum hall effect,''
  Phys.\ Rev.\ Lett.\  {\bf 62}, 82 (1988).
}

\lref\ZhangEU{
  S.~C.~Zhang,
  ``The Chern-Simons-Landau-Ginzburg theory of the fractional quantum Hall
  Int.\ J.\ Mod.\ Phys.\  B {\bf 6}, 25 (1992).
}

\lref\girvin{
Steven M. Girvin, ``The Quantum Hall Effect: Novel Excitations and Broken Symmetries", arXiv:cond-mat/9907002
}

\lref\DunneQY{
  G.~V.~Dunne,
  ``Aspects of Chern-Simons theory,''
  arXiv:hep-th/9902115.
}

\lref\KimBF{
  D.~K.~Kim and K.~S.~Soh,
  ``The effective action of (2+1)-dimensional QED: The effect of finite
  fermion density,''
  Phys.\ Rev.\  D {\bf 55}, 6218 (1997)
  [arXiv:hep-th/9606197].
}

\lref\HartnollAI{
  S.~A.~Hartnoll and P.~Kovtun,
  ``Hall conductivity from dyonic black holes,''
  Phys.\ Rev.\  D {\bf 76}, 066001 (2007)
  [arXiv:0704.1160 [hep-th]].
}

\lref\Shankar{
G.~Murthy and R.~Shankar,
``Hamiltonian theories of the fractional quantum Hall effect",
Rev. Mod. Phys, {\bf 75}, 1101 (2003)
}

\lref\DavisKA{
  J.~L.~Davis, M.~Gutperle, P.~Kraus and I.~Sachs,
  ``Stringy NJL and Gross-Neveu models at finite density and temperature,''
  JHEP {\bf 0710}, 049 (2007)
  [arXiv:0708.0589 [hep-th]].
}

\lref\GubserPX{
  S.~S.~Gubser,
  ``Breaking an Abelian gauge symmetry near a black hole horizon,''
  arXiv:0801.2977 [hep-th].
}

\lref\GubserZU{
  S.~S.~Gubser,
  ``Colorful horizons with charge in anti-de Sitter space,''
  arXiv:0803.3483 [hep-th].
}

\lref\NakanoXC{
  E.~Nakano and W.~Y.~Wen,
  ``Critical magnetic field in AdS/CFT superconductor,''
  arXiv:0804.3180 [hep-th].
}

\lref\AlbashEH{
  T.~Albash and C.~V.~Johnson,
  ``A Holographic Superconductor in an External Magnetic Field,''
  arXiv:0804.3466 [hep-th].
}

\lref\WenPB{
  W.~Y.~Wen,
  ``Inhomogeneous magnetic field in AdS/CFT superconductor,''
  arXiv:0805.1550 [hep-th].
}

\lref\GubserWV{
  S.~S.~Gubser and S.~S.~Pufu,
  ``The gravity dual of a p-wave superconductor,''
  arXiv:0805.2960 [hep-th].
}

\baselineskip15pt
\newsec{Introduction}

The AdS/CFT correspondence is a powerful tool for studying strongly coupled quantum
field theories.  Difficult field theory questions are recast in the language
of weakly coupled gravitational theories in one higher dimension, where they
are amenable to semi-classical analysis.     This ``geometrization" of non-perturbative field theory phenomena has already yielded much insight into supersymmetric gauge theories,  and recent efforts at describing the real-world
physics of quark-gluon plasmas and condensed matter systems show great promise.

\lref\prange{Richard E. Prange and Steven M. Girvin, eds, ``The Quantum Hall
  Effect" (2nd ed.)
(Springer-Verlag, 1990 }
\lref\das{Sankar Das Sarma and Aron Pinczuk, eds, ``Perspectives in Quantum Hall Effects" (John Wiley and Sons, 1997)}

In this paper we focus on the quantum Hall effect (QHE), and in particular its realization in AdS.   Discovered in the early 80s  and studied intensively ever since,
the integer and fractional quantum Hall effects arise in the rather exotic setting
of $2+1$ dimensional electron systems subjected to extremely low temperatures and
large magnetic fields. For reviews see, {\it e.g.} \refs{\prange,\das,\girvin,\Shankar}.   When immersed in a weak electric field, the  conductance of the system displays a striking series of plateaus.  On each plateau the ordinary conductance is zero, while the transverse (Hall) conductance is found, to startling
accuracy, to be a rational multiple of a ``fundamental" unit formed from the elementary constants of nature (Planck's constant,  the electron charge, and the speed of light).
Especially striking is the fact that these results exist even in the presence of finite temperature and disorder (within limits), as are of course present in any actual experimental setting.

\lref\HartnollAI{
  S.~A.~Hartnoll and P.~Kovtun,
  ``Hall conductivity from dyonic black holes,''
  Phys.\ Rev.\  D {\bf 76}, 066001 (2007)
  [arXiv:0704.1160 [hep-th]].
}

\lref\HartnollIH{
  S.~A.~Hartnoll, P.~K.~Kovtun, M.~Muller and S.~Sachdev,
  ``Theory of the Nernst effect near quantum phase transitions in condensed
  matter, and in dyonic black holes,''
  Phys.\ Rev.\  B {\bf 76}, 144502 (2007)
  [arXiv:0706.3215 [cond-mat.str-el]].
}

\lref\RobertsNS{
  M.~M.~Roberts and S.~A.~Hartnoll,
  ``Pseudogap and time reversal breaking in a holographic superconductor,''
  arXiv:0805.3898 [hep-th].
}

Being robust in the above sense, the Hall conductance can be thought
of as a topological quantity, and successful microscopic and
macroscopic accounts of the QHE incorporate this aspect in a crucial
way.   It also implies that the QHE is suitable for modelling via
the AdS/CFT correspondence.  In general, field theories dual to
weakly coupled AdS theories exist in some strongly coupled corner of
coupling space.    In favorable cases, one can relate the weakly
coupled field theory to its strongly coupled version through some
explicit extrapolation of the couplings.  Some observables are
either independent of the couplings or behave smoothly, and so a
weak-strong coupling comparison is meaningful. The quantum Hall conductance
is an example of such an observable. The classical (non-quantized)  Hall
conductance was given an AdS/CFT interpretation in
\refs{\HartnollAI}; see also \refs{\HartnollIH}.

\lref\weinberg{S. Weinberg, "Quantum Theory of Fields", Vol. II, Cambridge
University Press, Chapter 21.6. }
%

\lref\WilczekWY{
  F.~Wilczek,
  ``Quantum Mechanics Of Fractional Spin Particles,''
  Phys.\ Rev.\ Lett.\  {\bf 49}, 957 (1982).
}

\lref\ArovasYB{
  D.~P.~Arovas, J.~R.~Schrieffer, F.~Wilczek and A.~Zee,
  ``Statistical Mechanics Of Anyons,''
  Nucl.\ Phys.\  B {\bf 251}, 117 (1985).
}

Our AdS construction is motivated by effective field theory descriptions of the
QHE that exploit its relation to the BCS theory of superconductivity \refs{\ZhangWY,\ZhangEU}, along with
recent work on AdS versions of superconductors \refs{\GubserPX,\HartnollVX,\GubserZU,\NakanoXC,\AlbashEH,\WenPB,\GubserWV,\RobertsNS}.  To a particle theorist, the BCS
theory is of course nothing but the Higgs phenomenon corresponding to the spontaneous
breaking of electromagnetic $U(1)$ gauge invariance to a $\ZZ_2$ subgroup \weinberg. A simple AdS incarnation involves the condensation of a charged scalar field outside a black hole horizon.  We extend this construction by the introduction of an additional
gauge field with a nonzero theta term.  It plays the role of the ``statistical gauge
field" in the effective field theory description of the QHE.  The statistical gauge field
transmutes the electrons into bosons via Aharonov-Bohm phases \refs{\WilczekWY,\ArovasYB}, allowing the quantum
Hall fluid to be described in terms of Bose condensation.

The remainder of this paper is organized as follows.  In section 2 we give
an overview of some relevant facts concerning the QHE.  We discuss both model
independent aspects as well as the analogy with Bose condensation.  In section
3 we describe our AdS construction, and show that it indeed leads to the QHE.
Some further comments appear in section 4.

\lref\BrodieYZ{
  J.~H.~Brodie, L.~Susskind and N.~Toumbas,
  ``How Bob Laughlin tamed the giant graviton from Taub-NUT space,''
  JHEP {\bf 0102}, 003 (2001)
  [arXiv:hep-th/0010105].
}

\lref\BenaCS{
  I.~Bena and A.~Nudelman,
  ``On the stability of the quantum Hall soliton,''
  JHEP {\bf 0012}, 017 (2000)
  [arXiv:hep-th/0011155].
}

\lref\GubserDZ{
  S.~S.~Gubser and M.~Rangamani,
  ``D-brane dynamics and the quantum Hall effect,''
  JHEP {\bf 0105}, 041 (2001)
  [arXiv:hep-th/0012155].
}

\lref\SusskindFB{
  L.~Susskind,
  ``The quantum Hall fluid and non-commutative Chern Simons theory,''
  arXiv:hep-th/0101029.
}

\lref\BergmanQG{
  O.~Bergman, Y.~Okawa and J.~H.~Brodie,
  ``The stringy quantum Hall fluid,''
  JHEP {\bf 0111}, 019 (2001)
  [arXiv:hep-th/0107178].
}

\lref\HellermanYV{
  S.~Hellerman and L.~Susskind,
  ``Realizing the quantum Hall system in string theory,''
  arXiv:hep-th/0107200.
}

\lref\FreivogelVC{
  B.~Freivogel, L.~Susskind and N.~Toumbas,
  ``A two fluid description of the quantum Hall soliton,''
  arXiv:hep-th/0108076.
}

Previous work on the QHE in string theory, not directly within the AdS/CFT framework,
includes, \refs{\BrodieYZ,\BenaCS,\GubserDZ,\SusskindFB,\BergmanQG,\HellermanYV,\FreivogelVC}.

\newsec{Some Background on the Integer and Fractional Quantum Hall Effects}

In this section we give an overview of those aspects of the QHE most relevant to its description via the AdS/CFT correspondence.  While we make no claims to originality,
our presentation differs in some respects from other accounts in the literature.
In particular, we wish to emphasize the generality of the  QHE, in the sense that
it relies on only a few underlying assumptions that we spell out.  Whether these
assumptions are met or not is a question that can only be answered in terms
of a specific microscopic realization.

\subsec{Model independent considerations}

Consider some system of charged particles in $2+1$ dimensions.  These particles
could be fermions or bosons, or both could be present simultaneously; for convenience we will
simply call them electrons.   The electrons are allowed
to interact with each other as well as with any external fields that are present.
The theory can be either Galilean or Lorentz invariant.   We assume that over the
largest length scales the Hamiltonian is invariant under time and space translations as well as spatial rotations, but not parity.\foot{Recall that parity in $2+1$ dimensions is defined
as sign reversal of one of the spatial directions, since flipping both spatial
directions is equivalent to a rotation.}

The quantum Hall effect arises if the particles are in a state with an energy gap; that is to say, if the system is in an energy eigenstate $|E_0\rangle$, the Hamiltonian
has no eigenvalues between $E_0$ and $E_0+\Delta$, where $\Delta$ is a positive number.

To see that this leads to the QHE we proceed as follows.  Let $A_\mu$ be the
electromagnetic gauge field, and let $\overline{A}_\mu$ be its expectation value (in some gauge)  in the gapped state of interest. In the standard QHE $\overline{A}_\mu$
corresponds to a constant magnetic field.   Allow small fluctuations by writing
$A_\mu = \overline{A}_\mu + a_\mu$.   We can derive an effective action for $a_\mu$
by performing the path integral over the fluctuations of the electrons.  Because
the electrons have an energy gap, this effective action admits an expansion in
terms of local operators, with low dimension operators dominating on large length
scales.    Since this action must be gauge invariant, there are no terms with
zero derivatives, while at single derivative order the only possibility is a
Chern-Simons term,\foot{Conventions: our metric signature is $(-,+,+)$ and  we choose orientation $\epsilon_{012}=1$. }
\eqn\aa{ S(a) =  {k\over 4\pi} \int\! d^3x~ \epsilon^{\alpha \beta \gamma}a_\alpha \p_\beta a_\gamma +  \dots }
where we have indicated that additional terms in the action have two or more derivatives.  $k$ is a pure number when we work  in units such that $\hbar=c=|e|=1$.
The Chern-Simons density famously varies by a total derivative under gauge transformation, and so the action is gauge invariant.  The action
is parity odd, and so nonzero $k$ requires parity non-invariance of the underlying
electron system.  Parity can be violated by $\overline{A}_\mu$, by inter-electron interactions, or otherwise.

We can now compute the response of the system to an applied  field.
The induced current is
\eqn\ab{ j^\alpha = {\delta S \over \delta a_\alpha} = {k\over 2\pi}  \epsilon^{\alpha\beta\gamma} \p_\beta a_\gamma + \ldots }
For a constant electric field, $E_i = \p_0 a_i - \p_i a_0$, the current is thus
\eqn\ac{ j^i = {k\over 2\pi}  \epsilon^{ij}E_j~,}
which identifies the  conductance as
\eqn\ad{ \sigma_{ij} = {k\over 2\pi}  \epsilon_{ij}~.}
In particular, the longitudinal conductance vanishes, while the transverse
conductance is fixed by $k$.

In the standard experimental realization of the QHE,
the external magnetic field
is varied at fixed charge density.  To explain the observed plateaus we have
to explain why $k$ does not vary along with the magnetic field.  Further,
since $k$ is observed to be a rational number, it is apparently insensitive
to  much of the detailed structure of the Hamiltonian, a point which also
requires explanation.

The explanation follows from a ``non-renormalization theorem" for $k$.
In particular, let the action of the system depend on some adjustable parameters, denoted collectively by $\alpha$.  We require that the action remain gauge invariant as we vary the $\alpha$, and in particular, we require
this to be the case even when we allow these parameters to be spacetime dependent: $\alpha= \alpha(x^\mu)$.    Examples of such parameters include the
strength of the external magnetic field and the gauge coupling.\foot{For the latter, normalize the gauge field so the gauge coupling does not appear in the
gauge transformation law; {\it i.e.}, so that the Maxwell term is $-{1\over 4}\int\! d^3 x ~{1\over e^2(x)} F^{\mu\nu}F_{\mu\nu}$.}

Starting from our original action, we consider a family of actions $S(\alpha)$ depending on {\it constant} $\alpha$ parameters.
We stipulate that the gap $\Delta$ remain finite within this family.  Generically, this will be the case for some open neighborhood in $\alpha$ space.\foot{ But not always, such as in a system without any  disorder, as our next Dirac fermion example illustrates. }   For each member of this family we can integrate out the electron fluctuations to obtain an effective action as in \aa.  The non-renormalization theorem states that $k$ is independent of $\alpha$.

The proof is extremely simple, and parallels the proof of the non-renormalization of the Fayet-Iliopoulos parameter in supersymmetric gauge theories \refs{\DineUI,\WeinbergUV,\SakamotoCX}.   Consider some slowly varying $\alpha$ parameters.  Then, instead of
\aa\ we will get
\eqn\ae{ S(a,\alpha) =   {1\over 4\pi} \int\! d^3x~ k[\alpha(x)]\epsilon^{\alpha \beta \gamma}a_\alpha \p_\beta a_\gamma +  \dots }
But this term is not gauge invariant for nonconstant $\alpha$, and it is
easy to check that its gauge variation cannot be canceled by any other local term.
Since the action is assumed to be gauge invariant, the only possibility is
that $k$ is independent of $\alpha$, completing the proof.  We stress the crucial role played by the condition that the energy gap remains finite; if
$\alpha$ is varied such that the gap disappears, $k$ can change.  In the
QHE this is precisely what happens as we transition from one plateau
to another.

In the case that $\alpha$ refers to the gauge coupling $e$, the theorem states
that $k$ receives corrections at one-loop, but not beyond.  This follows
since when we normalize the Maxwell term to $ -{F^2 \over 4e^2}$ the
$l$-loop term carries the $e$ dependence $e^{2l-2}$, and so only $l=1$
gives an $e$ independent result.

\lref\SemenoffDQ{
  G.~W.~Semenoff,
  ``Condensed Matter Simulation Of A Three-Dimensional Anomaly,''
  Phys.\ Rev.\ Lett.\  {\bf 53}, 2449 (1984).
}

\lref\NovoselovKJ{
  K.~S.~Novoselov {\it et al.},
  ``Two-Dimensional Gas of Massless Dirac Fermions in Graphene,''
  Nature {\bf 438}, 197 (2005)
  [arXiv:cond-mat/0509330].
}

\lref\Zhangetal{
  Y. Zhang {\it et al.},
  ``Experimental observation of the quantum Hall effect and Berry's phase in graphene''
  Nature {\bf 438}, 201 (2005)
}

As a concrete example that will be useful in what follows, consider the
case of Dirac fermions in a constant magnetic field,
\eqn\af{ S_\psi = \int\!d^3x\,\psibar (i \p\!\!\!\slash + A \!\!\!\slash +m+\mu \gamma^0) \psi~, }
where we have included a chemical potential $\mu$ to control the
charge density. The massless limit of this theory is relevant for
the recently observed anomalous integer quantum Hall effect in
graphene \refs{\SemenoffDQ,\NovoselovKJ,\Zhangetal}.

Parity in $x^1$ acts on the Dirac field as  $\psi(x^1) \rightarrow \gamma^1 \psi(-x^1)$,
which implies that the mass term is parity odd:  $m\psibar \psi
\rightarrow -m\psibar \psi$.

Writing $A= \overline{A}+a$, where $\overline{A}$ describes a constant
magnetic field $B$, computation of  the one-loop effective action for $a$ yields a Chern-Simons term with coefficient \KimBF\
\eqn\ag{\eqalign{ k &= {1\over 2} \Bigg\{- {\rm sign}(m)\Theta(m^2 -\mu^2) \cr &\quad\quad +2 {\rm sign}(\mu) \Theta(\mu^2-m^2) \left[{\mu^2-m^2 \over 2  |B|  }+\sum_{n=1}^\infty {1 \over \pi n}\sin \left(\pi n {(\mu^2-m^2) \over  B}\right)\right]\Bigg\}~. }}

To interpret this, we note that solutions of the Dirac equation
$ (i \p\!\!\!\slash + \overline{A} \!\!\!\slash +m)\psi =0$ have
energy spectrum
\eqn\ah{ \omega_n = \sqrt{m^2 +2nB}}
where $n=0,1,2,\ldots$ or $n=1,2,\ldots$ depending on whether the
electron spin is parallel or anti-parallel to the magnetic field.
For orientation, note that in the non-relativistic limit $m\gg B$, we have $\omega_n \approx m -{1\over 2} {B\over m} + (n+{1 \over 2}){B\over m}$, which is the familiar Landau level spectrum plus a zero point energy.   As in the
non-relativistic case, the degeneracy for each spin state is $BA/(2\pi)$, where
$A$ is the area of the system.

The system has an energy gap when a given energy level is completely filled,
which happens for chemical potential
\eqn\ai{\mu^2 = m^2 +2pB~,\quad  p =1, 2, \ldots .}
This gives Chern-Simons coupling
\eqn\aj{ k =  p ~{\rm sign}(B)~.}
This illustrates the content of the non-renormalization theorem: for these
gapped states the Chern-Simons coupling is independent of the magnitude of $m$ and $B$ (at fixed $\mu$).   On the other hand, for generic $\mu$, we see that $k$ does depend on these parameters.

To fully explain the integer QHE we need one final ingredient.   The above
system does not exhibit plateaus as we vary $B$ keeping everything else fixed.
The reason is that if we start with fully filled energy levels and then
change $B$, we inevitably end up with partially filled levels and hence no
energy gap.  The integer QHE only occurs if there are additional spatially localized
states in the spectrum, with energies between those of the Landau levels.
These states arise from disorder in the material.  Then as we vary the magnetic
field we simply fill up these localized states.  Being localized, these
states cannot affect the Chern-Simons coupling, and hence the Hall conductivity
remains fixed until we reach the next Landau level.   At this point $k$ jumps
to a new value, and the process repeats itself.

Although we explicitly computed the Chern-Simons coupling only  for free electrons, the non-renormalization theorem tells us that we can turn on inter-electron
interactions without changing the result (provided we do not destroy the gap).
Therefore, our free fermion derivation of the integer QHE actually applies
to a whole ``universality class" of theories that can be smoothly connected
to the free theory.  This explains the robust nature of the integer QHE.

These considerations highlight that the observed {\it fractional} QHE plateaus
must correspond to gapped systems that cannot be smoothly deformed to
noninteracting electrons.  Instead, they correspond to new universality
classes of interacting electrons.  Just based on our general considerations,
there is no way of saying which universality classes, {\it i.e.} which
values of $k$, can actually occur.  But if a gapped state with a given $k$ does occur, we now understand why there is a quantum Hall plateau in the conductance,
independent of the microscopic details of the system.

\lref\LaughlinFY{
  R.~B.~Laughlin,
  ``Anomalous quantum Hall effect: An incompressible quantum fluid with
  fractionally charged excitations,''
  Phys.\ Rev.\ Lett.\  {\bf 50}, 1395 (1983).
}

\subsec{Zhang-Hansson-Kivelson model of the fractional QHE}

The most prominent fractional quantum Hall plateaus are those for
$k= {1\over 2p-1}$, with $p=1,2, \ldots$.   The existence of these
states was originally explained by Laughlin \LaughlinFY\ in terms of
an explicit class of electron wavefunctions. Alternatively, one can
use the language of effective field theory to model the
long-distance aspects of the problem.  There are a number of
different effective field theories on the market (for a review of
some of these, see \Shankar).  Here we will follow the approach in
\ZhangWY\ (ZHK), since its connection with superconductivity suggests
a natural AdS implementation. It would be interesting to find
implementations also for other effective theories.

 The ZHK Lagrangian is
\eqn\zza{\eqalign{ {\cal L} &= {k\over 4\pi} \epsilon^{\mu\nu\rho}a_\mu \p_\nu a_\rho +i\psi^* (\p_0-i(a_0+A_0))\psi
-{1\over 2m}|\left(\p_i-i(a_i+A_i)\right)\psi|^2 \cr &\quad\quad -{1\over 2} \int\! d^2x' |\psi(x)|^2 V(x-x')
|\psi(x')|^2 ~.}}
Here $\psi$ is a complex {\it bosonic} field; $A_\mu$ is the electromagnetic gauge field, treated here as a non-dynamical external field; and $a_\mu$ is an additional ``statistical gauge field".
The role of the statistical gauge field is to transmute the bosons into
fermions via Aharonov-Bohm phases \refs{\WilczekWY,\ArovasYB}.  This occurs provided we take
\eqn\zzb{ k ={1 \over 2p-1}~,\quad  p=1, 2, \ldots ~.}
For such values of $k$, the theory is equivalent to a system of fermions
minimally coupled to the electromagnetic field and interacting via the
potential $V(x-x')$.  The advantage of the bosonic representation is that it
allows for a description of the gapped states in terms of a classical Higgs mechanism.

We now sketch how this model accounts for the fractional QHE plateaus.
We will be somewhat schematic, since the explicit computations are
precisely parallel to those in the AdS construction that follows.  The idea
is to first look for homogeneous solutions, representing a constant charge
density in a constant magnetic field. These solutions break the electromagnetic gauge symmetry, yielding a gapped spectrum.  Such solutions exist only at filling fraction $\nu=k$.   The excitations at this filling fraction include
vortices, which acquire fractional charges via the Chern-Simons interaction.
Varying the filling fraction away from $\nu=k$, the state of the system is described by a gas of vortices.   The gap persists, and so the Hall conductivity is pinned at the value ${k\over 2\pi}$.

In somewhat more detail, the first step is to consider the $a_0$ equation of motion, which ties the statistical magnetic $b$ field to the particle density:
\eqn\zzc{ {k  \over 2\pi}b  = -|\psi|^2 = -\rho~. }
The $a_i$ equation forces $b=-B$, and hence the filling fraction is determined:
\eqn\zzca{ {k  \over 2\pi}B  =\rho \quad \Rightarrow \quad \nu = {2\pi \rho \over B} = k~. }

The conductance at arbitrary filling fraction is computed as follows.
We can split up the action as
\eqn\zzg{ S = S_{CS}(a) + S_\psi(\psi, a+A)~,}
and then compute the current in a constant external electric field as
\eqn\zzcc{ j^i ={\delta S_{\psi} \over \delta A_i} = {\delta S_{\psi} \over \delta a_i}=-{\delta S_{CS} \over \delta a_i}=-{k\over 2\pi} \epsilon^{ij}f_{0j}~.}
Since the total gauge field $A+a$ is massive, we have $f_{0j}=-F_{0j}$ for
constant electric fields,  and hence the Hall conductance is indeed ${k\over 2\pi}$.

Reviewing the chain of reasoning, it becomes apparent that the conclusions
are insensitive to the detailed structure of the action, a point that will
be important in the AdS version.  We could have started from a general action of
the form $S_{CS}(a)+S_\psi(\psi, A+a)$.  As long as $\psi$ condenses to
break the gauge symmetry, we generically deduce the existence of fractional
QHE plateaus with Hall conductance ${k\over 2\pi}$.

\lref\HertogRR{
  T.~Hertog,
  ``Towards a novel no-hair theorem for black holes,''
  Phys.\ Rev.\  D {\bf 74}, 084008 (2006)
  [arXiv:gr-qc/0608075].
}

\lref\BalasubramanianSN{
  V.~Balasubramanian, P.~Kraus and A.~E.~Lawrence,
  ``Bulk vs. boundary dynamics in anti-de Sitter spacetime,''
  Phys.\ Rev.\  D {\bf 59}, 046003 (1999)
  [arXiv:hep-th/9805171].
  V.~Balasubramanian, P.~Kraus, A.~E.~Lawrence and S.~P.~Trivedi,
  ``Holographic probes of anti-de Sitter space-times,''
  Phys.\ Rev.\  D {\bf 59}, 104021 (1999)
  [arXiv:hep-th/9808017].
}

\newsec{AdS Construction}

We now turn to the gravitational description of the QHE.  We first review
a recent model of an AdS$_4$ superconductor, and then show how to adapt
it to the case of the QHE.

\subsec{AdS superconductor}

The authors of \HartnollVX\ consider a planar, asymptotically AdS$_4$ black hole,
\eqn\ba{ ds^2 = -f(r)dt^2+{dr^2 \over f(r)} +r^2\left((dx^1)^2+ (dx^2)^2\right)~,}
with
\eqn\bb{f(r) = {r^2\over L^2 }-{M\over r}~.}
This is dual to a $2+1$ dimensional quantum field theory at the Hawking temperature,\foot{Taking the temperature
strictly to zero yields a singularity in the solutions that follow.}
\eqn\bc{ T={3M^{1/ 3} \over 4\pi L^{4/3}}~.}
Adding a gauge field in the bulk with a gauge invariant action is equivalent
to considering a boundary theory with a $U(1)$ global symmetry.  If this symmetry is spontaneously broken, then the theory will exhibit superconductivity with respect to external gauge fields coupled to the $U(1)$ current.
To spontaneously break the symmetry we would like some charged operator
to acquire an expectation value.  Via the AdS/CFT correspondence, this means
that we need to  find a black hole solution with hair.  The simplest possibility is to consider a charged scalar field $\psi$ with a normalizable profile.

No-hair theorems/conjectures impose some restrictions on the possible hair
that can arise \HertogRR.  The specific example considered in \HartnollVX\  consists of the bulk Lagrangian
\eqn\bd{ {\cal L} = -{1 \over 4}F^{MN}F_{MN} - V(\psi) -|\p \psi -iA \psi|^2~,}
with
\eqn\bez{V(\psi) = -2 {|\psi|^2  \over L^2}~.}
Working in a small amplitude limit such that backreaction on the metric
can be neglected, the authors find a solution with nonzero $A_t(r)$ and $\psi(r)$ with large $r$ behavior
\eqn\bfz{ A_t(r) = \mu -{\rho \over r} +\ldots~,\quad \psi(r) = {\psi^{(1)}\over r}+{\psi^{(2)}\over r^2} + \ldots ~.}
After choosing boundary conditions such that either $\psi^{(1)}$ or $\psi^{(2)}$ vanishes,\foot{Both choices yield normalizable solutions.} there is a one-parameter family of solutions that we can parameterize by the charge density $\rho$.   The solution represents a charged black hole with scalar hair, and is dual to a field theory in a state with a spontaneously broken $U(1)$
global symmetry.  The size of the condensate is proportional to amplitude of the normalizable scalar mode $\psi^{(1,2)}$ \refs{\BalasubramanianSN}.

To study the response of the system to applied fields we can look for more
general solutions in which the gauge field obeys the boundary condition
\eqn\bg{ A_\mu(x^\mu,r) = A_\mu^{(0)}(x^\mu) + {1\over r} A_\mu^{(1)}(x^\mu) +\ldots~.}
Here $x^\mu$ denote the coordinates of the $2+1$ boundary, and we have chosen the gauge $A_r=0$.  The Dirichlet problem consists of solving the field equations for prescribed $A_\mu^{(0)}(x^\mu)$.  The standard AdS/CFT
dictionary tells us that the resulting on-shell action is equal to the partition function of the boundary theory in the presence of external sources $A_\mu^{(0)}(x^\mu)$ coupled to the $U(1)$ current.

Since the gauge symmetry in the bulk is spontaneously broken, the on-shell
action admits a derivative expansion.  At quadratic order in fluctuations
around the background solution, this action will take the form
\eqn\bh{S = \int\! d^3x ~ \pi^{\mu\nu} A_\mu^{(0)} A_{\nu}^{(0)} + {\rm derivative~terms} ~,}
with nonzero $\pi^{00}$ and $\pi^{11}=\pi^{22}$.  The values of $\pi^{\mu\nu}$, which are a function of the charge density $\rho$, can be extracted from the
numerical results in \HartnollVX, but the precise values will not be needed.
What really matters is that $\pi^{\mu\nu}$ is nonzero, which  is in turn
a direct consequence of the Higgs mechanism in the bulk.  On very general
grounds, the existence of the mass term in \bh\ implies superconductivity \weinberg, and also plays a key role in giving fractional quantum Hall behavior.

\subsec{ Quantum Hall construction}

We are now ready to describe our bulk model of the QHE.  The background geometry is
the same as in \ba, and the action is similar to \bd, but with the addition
of an extra gauge field $a_M$ with a nonzero theta term,
\eqn\ca{ {\cal L} = {k\over 4\pi} \epsilon^{MNPQ}f_{MN}f_{PQ}-{1 \over 4}(F+f)^{MN}(F+f)_{MN} - V(\psi) -|\p \psi -i(A+a) \psi|^2~.}
The theta term is equivalent to a Chern-Simons term on the boundary. In order
for this to be a dynamical Chern-Simons term we allow a asymptotic behavior of
$a_\mu$ as in \bg, but now also let $a_\mu^{(0)}$ freely vary.  That is, we demand that the action be stationary under variations of  $a_\mu^{(0)}$.  By
contrast, we hold $A_\mu^{(0)}$ fixed, so that it can be interpreted as the
external electromagnetic field.

At this effective field theory level $k$ is an arbitrary number.  In the
ZHK model the values \zzb\ were singled out in order for the theory to describe
fermions of unit charge, {\it i.e.} electrons.   In the present context, we expect that
if our model has some underlying weak-coupling brane description then certain
special values of $k$ will emerge.  But for now, $k$ is arbitrary.

It will often be convenient to write the total action as
\eqn\cab{ S= S_{CS}(a) + S_\psi(\psi,A+a)~,}
where $S_{CS}(a)$ denotes the first term in \ca, and $S_\psi$ is everything else.

The $a_t^{(0)}$ equation of motion relates the charge density to the statistical magnetic field,
\eqn\cb{  - {k\over 2\pi} b={\delta S_{CS} \over \delta a^{(0)}_t} =  -{\delta S_\psi \over \delta
a^{(0)}_t} = -{\delta S_\psi \over \delta
A^{(0)}_t}= -\rho~.}

Next, we look for a solution of the bulk equations of motion with a constant
magnetic field on the boundary.   Since the theta-term does not contribute
to the $A$ equation of motion, the combination $A+a$ obeys the same bulk equation as did $A$ in the superconductor
setup.  Therefore, we can use the same solution as before, but now for $A+a$:
\eqn\cc{ A_t(r)+a_t(r) = \mu -{\rho \over r} +\ldots~,\quad \psi(r) = {\psi^{(1)}\over r}+{\psi^{(2)}\over r^2} + \ldots ~,}
and we can trivially turn on constant magnetic fields as well,
\eqn\cd{ B=-b ={\rm constant}~.}
Any magnetic field is allowed by the bulk equations, but this freedom is removed by \cb, which fixes
\eqn\ce{ B = - b = {2\pi \rho \over k}~.}
This equation determines the preferred filling fractions for which these simple homogeneous solutions exist.

Next we compute the conductance by turning on a constant electric field in
$A^{(0)}_\mu$.  The current is given by differentiating the on-shell action,
\eqn\cf{ j^\mu ={\delta S \over \delta  A^{(0)}_\mu}~.}
Given the structure \cab, and the $a_\mu^{(0)}$ equation of motion, we have
\eqn\cg{ j^\mu ={\delta S_\psi \over \delta  A^{(0)}_\mu} ={\delta S_\psi \over \delta  a^{(0)}_\mu} = -{\delta S_{CS} \over \delta  a^{(0)}_\mu} = {k\over 2\pi} \epsilon^{\mu\nu\rho} \p_\nu a_\rho^{(0)}   ~.}

To complete the computation we need to express $a_\mu^{(0)}$ in terms of
$A_\mu^{(0)}$.   For this we use the derivative expansion of the on-shell
action, keeping in mind that we are interested in the response to a {\it constant} electric field.  At quadratic order in fluctuations, zero and
one derivative terms are given by combining \bh\ and the Chern-Simons term,
\eqn\ch{ S = \int\! d^3 x~\left( \pi^{\mu\nu} (A+a)^{(0)}_\mu (A+a)^{(0)}_\nu + {k\over 4\pi} \epsilon^{\mu\nu\rho} a_\mu^{(0)} \p_\nu a_\rho^{(0)} +\ldots \right)~.}
The $a_\mu^{(0)}$ equation of motion then implies
\eqn\ci{ a_\mu^{(0)}= -A_\mu^{(0)} + \ldots }
For a constant electric field, the exact current is therefore  given by
\eqn\cj{ j^i = {k\over 2\pi} \epsilon^{ij} F_{0j}   ~,}
yielding the conductance
\eqn\ck{ \sigma_{ij} = {k\over 2\pi} \epsilon_{ij}~.}

As shown above, homogenous solutions exist only for magnetic field
$B={2\pi \rho \over k}$.  What happens if we vary $B$ while keeping
the total charge fixed, as is done in a real  experiment? The only
possibility is that the solution becomes inhomogeneous, and we need these
inhomogeneities to carry charge.   There is a natural
candidate for these.  Instead of taking
constant $\psi^{(1,2)}$ we can look for vortex configurations.  These
vortices will be accompanied by magnetic $b$ flux, in order to render
the energy finite, just like in the Abelian Higgs model.  The Chern-Simons
interaction implies that a vortex carrying $r$ units of magnetic flux acquires
a charge\foot{In the condensed matter context, this result is sometimes used to argue that $k$ must be rational so that integer charge objects appear in the spectrum.}
\eqn\cl{ Q = kr~.}
We expect  that the solution for arbitrary magnetic field consists of
a distribution of vortices living in the homogeneous background, where the
latter continues to obey $B={2\pi \rho \over k}$.   This is the
effective field theory version of Laughlin's quasiparticle explanation of
the fractional QHE states \LaughlinFY.

In order that these vortices not contribute to the conductance they need
to be spatially localized.   In the condensed matter context, this occurs
through the presence of disorder; the vortices bind to impurities in the
sample.  This effect is not taken into account explicitly in the simple ZHK Lagrangian \zza, but can be added by hand.  Similarly, here we could model
the effects of disorder by generalizing the boundary conditions to include
spatial variation, which will localize the vortices.

\newsec{Discussion}

This work is a first step in studying the QHE within  the AdS/CFT correspondence.  We have only given an AdS version of the story, and
to exploit the real power of AdS/CFT we would like to be able to compare
against some dual non-gravitational description, perhaps in terms of
intersecting branes.  Also, on the AdS side we just studied a toy
theory, not obtained directly from string theory.  However, the basic
mechanism is quite general, and so will apply to a large class of theories.

There are many other things to study within this model, such as
the full AC conductivities, vortex solutions, edge states, multilayer systems, etc.  It would also be desirable to generalize the model to allow for transitions among different plateaus.  This would involve promoting $k$ to
a dynamical variable. We hope to return to these problems in the future.

\bigskip
\noindent {\bf Acknowledgments:} \medskip \noindent  We thank
N. Jokela, and S. Nowling for discussions. The work of PK is
supported in part by the NSF grant PHY-0456200.

\listrefs
\end

\newsec{Internal note on conventions}

4D Metric signature is $(-,+,+,+,+)$.    Choose $\epsilon_{012r}=\epsilon_{012}=1$.

Hall current formula is $j^i = {k\over 2\pi}\epsilon^{ij}F_{0j}$.  Current is defined in terms of action as $j^\mu = {\delta S \over \delta A_\mu}$.  Hall
current then comes from a CS term
\eqn\ya{ S_{CS} = {k\over 4\pi}\int\!d^3x ~ \epsilon^{\mu\nu\rho}A_\mu\p_\nu A_\rho }
Also have
\eqn\yb{k=\nu = {2\pi n \over B}={\rm filling~fraction}}
